# Lesion characterization in spectral photon-counting tomosynthesis

Björn Cederström[a], Karl Berggren[a,b], Klaus Erhard[c], Mats Danielsson[b], Erik Fredenberg[a], and Matthew Wallis[d]

[a]Mammography Solutions, Philips, Torshamnsg. 30A, 164 40 Kista, Sweden
[b]Dept. of Physics, Royal Inst. of Technology (KTH), 106 91 Stockholm, Sweden
[c]Philips Research Laboratories, Röntgenstrasse 24–26, 22335 Hamburg, Germany
[d]Cambridge Breast Unit, Addenbrooke's Hospital, Cambridge, United Kingdom

## ABSTRACT

It has previously been shown that 2D spectral mammography can be used to discriminate between (likely benign) cystic and (potentially malignant) solid lesions in order to reduce unnecessary recalls in mammography. One limitation of the technique is, however, that the composition of overlapping tissue needs to be interpolated from a region surrounding the lesion, and this uncertainty could potentially be reduced using the 3D information from spectral tomosynthesis. We present a first investiagtion of such an application. A phantom experiment was designed to simulate a cyst and a tumor, where the tumor was overlaid with a structure that made it mimic a cyst. In 2D, the two targets appeared similar in composition, whereas spectral tomosynthesis revealed the exact compositional difference. However, the loss of discrimination signal due to spread out from the plane of interest was of the same strength as the reduction of anatomical noise. A test on clinical tomosynthesis images of solid lesions was inconclusive and more data, as well as refinement of the calibration and algorithm, are needed.

## 1. INTRODUCTION

Solitary well defined mass lesions are a common mammographic finding, which contributes approximately 20% of overall recalls at screening.[1] A large number of these lesions are found to be simple cysts when assessed with ultrasound and do not require further clinical evaluation; cancer rates in solid probably benign lesions are less than 2%. Improving lesion characterization at screening and thereby lowering the number of recalls for simple cysts would be desirable to reduce both the costs of the screening program as well as patient anxiety. Spectral x-ray imaging is an emerging technology that measures the energy dependence of x-ray attenuation. One potential application of spectral imaging is to characterize breast lesions identified in mammographic screening with the aim to reduce the number of recalls for cysts.[2] We have previously shown in specimen experiments that it is feasible to discriminate between cyst fluid[3] and solid tissue[4] using 2D spectral mammography. These results motivated a clinical pilot study for discriminating between cystic and solid lesions,[1] which demonstrated a specificity (correctly classified cysts) of about 50% at the 99% sensitivity (correctly classified solids) level. Even though these results are encouraging, there is room for improvement. One limitation of lesion characterization in 2D is that the breast composition above and below the lesion needs to be estimated by interpolation from a region surrounding the lesion, assuming that there are no strong local changes in the composition.

The aim of this investigation was to (1) demonstrate that lesion characterization can be done with spectral tomosynthesis, and (2) investigate whether the 3D information can reduce the uncertainty from the interpolation of surrounding tissue. It should be acknowledged, though, that tomosynthesis, being limited in the angular span, is not a quantitative technique like computed tomography, and that e.g. uncertainty of the local breast thickness will play an important role also in this case.

During previous studies of 2D lesion characterization, a number of potentially limiting factors have been identified:[5]

Send correspondence to Björn Cederström. E-mail: bjorn.cederstrom@philips.com

- Uncertainty in tissue composition – This factor will remain the same in tomosynthesis. Note though that the clinical results in Ref.[1] indicate that some of the spread in tumor tissue composition reported in Ref.[4] is likely due to sample preparation effects.
- Statistical noise – On a global scale this will remain the same in tomosynthesis, since the dose will be the same. However, local estimates in the 3D volume may suffer more from statistical noise than local estimates in the projection images.
- Calibration imperfections – This should ideally be the same for 2D and tomosynthesis, but may differ due to different implementation of calibration procedures. This is the first 3D implementation and the algorithm may need refinement.
- Errors in interpolation of breast glandularity (structural overlap) – Here there is potential for improvement in tomosynthesis, with less structural overlap, better local estimation of glandularity, and a smaller region in which interpolation from the surrounding is necessary.
- Errors in interpolation of breast thickness – Due to the limited tomographic angle, tomosynthesis does not necessarily have any benefit here.

## 2. MATERIALS AND METHODS

### 2.1 Spectral photon-counting tomosynthesis clinical prototype

The Philips MicroDose S0 spectral tomosynthesis system (Philips Mammography Solutions, Solna, Sweden) is a not commercailly available clinical prototype. The system comprises an x-ray tube, a pre-collimator, and an image receptor, all mounted on a rigid arm (1, left). The image receptor consists of photon-counting silicon strip detectors with corresponding slits in the pre-collimator (1, right). To acquire an image, the arm is rotated around a point below the patient support so that the detector is scanned across the object. Each detector line views each point in the object from a unique angle, which results in a data set that can be used for 3D reconstruction. The width of the detector and the ppoint of rotation-to-detector distance yield a tomographic angle of about 11$^{o}$.

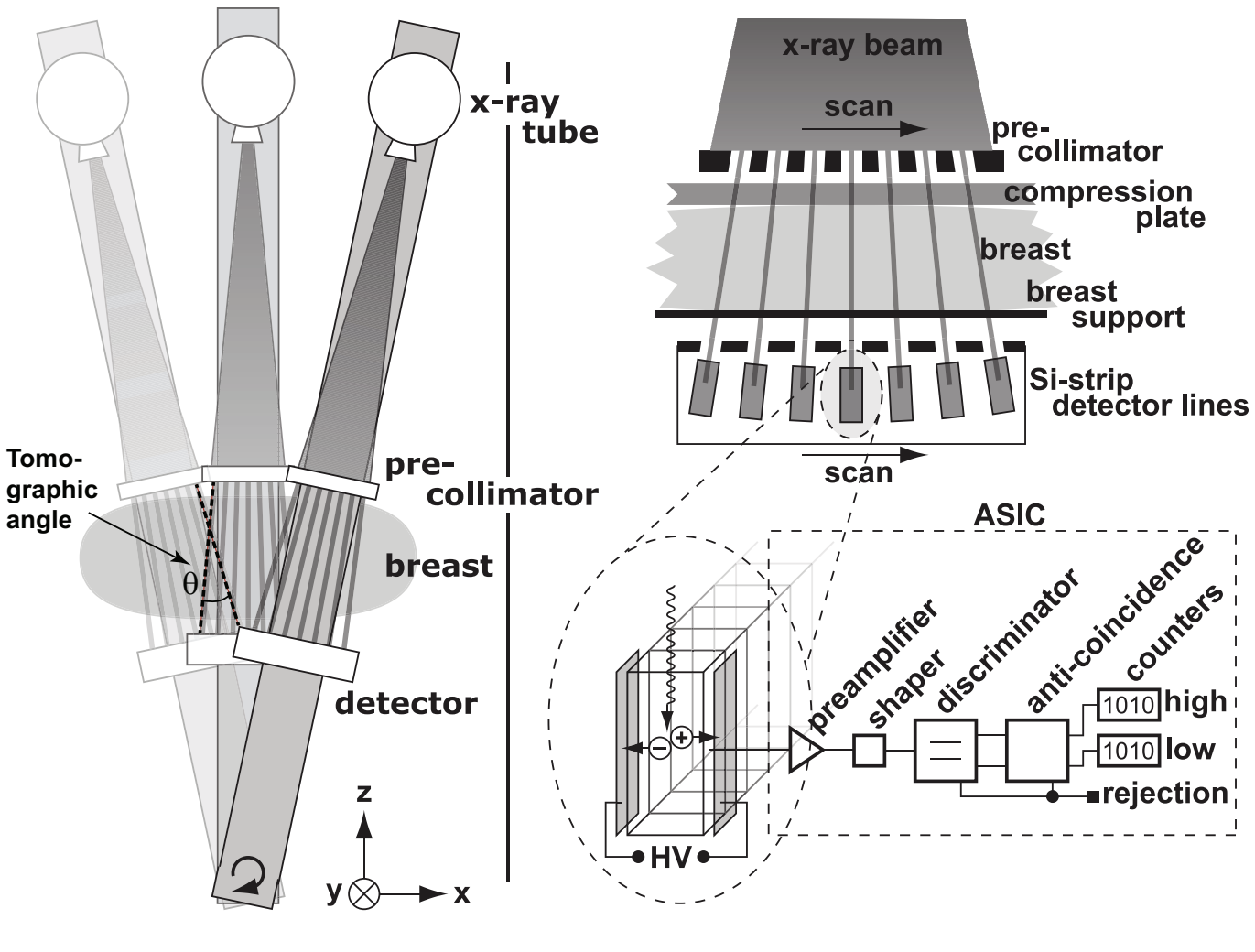

Figure 1. **Left:** Schematic of the Philips MicroDose S0 prototype tomosynthesis system. Note the indication of the tomographic angle, which is the angle at which the outermost rays cross at a certain point in the image volume. **Right:** The photon-counting spectral image receptor and electronics.

Photons that interact in the detector are converted to pulses with amplitude proportional to the photon energy. Virtually all pulses below a few keV are generated by noise and are therefore rejected by a low-energy threshold. A high-energy threshold sorts the detected pulses into two bins according to energy, which enables spectral imaging.

## 2.2 Calibration, reconstruction and material decomposition

See Fig. 2 for a schematic of the various calibration and reconstruction steps. The data from the two energy bins (*high* energy bin, and *sum* of low and high energy bins) are first flat-field calibrated independently. The raw photon counts are mapped to equivalent PMMA thickness using a look-up table from calibration using a PMMA step-wedge phantom. Each detector channel and energy bin is calibrated individually. This calibration step is performed on a regular basis to account for effects such as x-ray tube drift and detector and electronics temperature variations. Each energy bin is then separately iteratively reconstructed to image stacks expressed in PMMA thickness. The reconstruction algorithm (SART) was modified to yield a globally linear behavior. The images were binned to 0.4 x 0.4 x 2 $mm^3$ voxel size, in order to reduce noise and since fine-scale structures are unimportant for lesion characterization.

The next step is the material decomposotion into two basis materials, aluminum (Al) and polyethylene (PE).

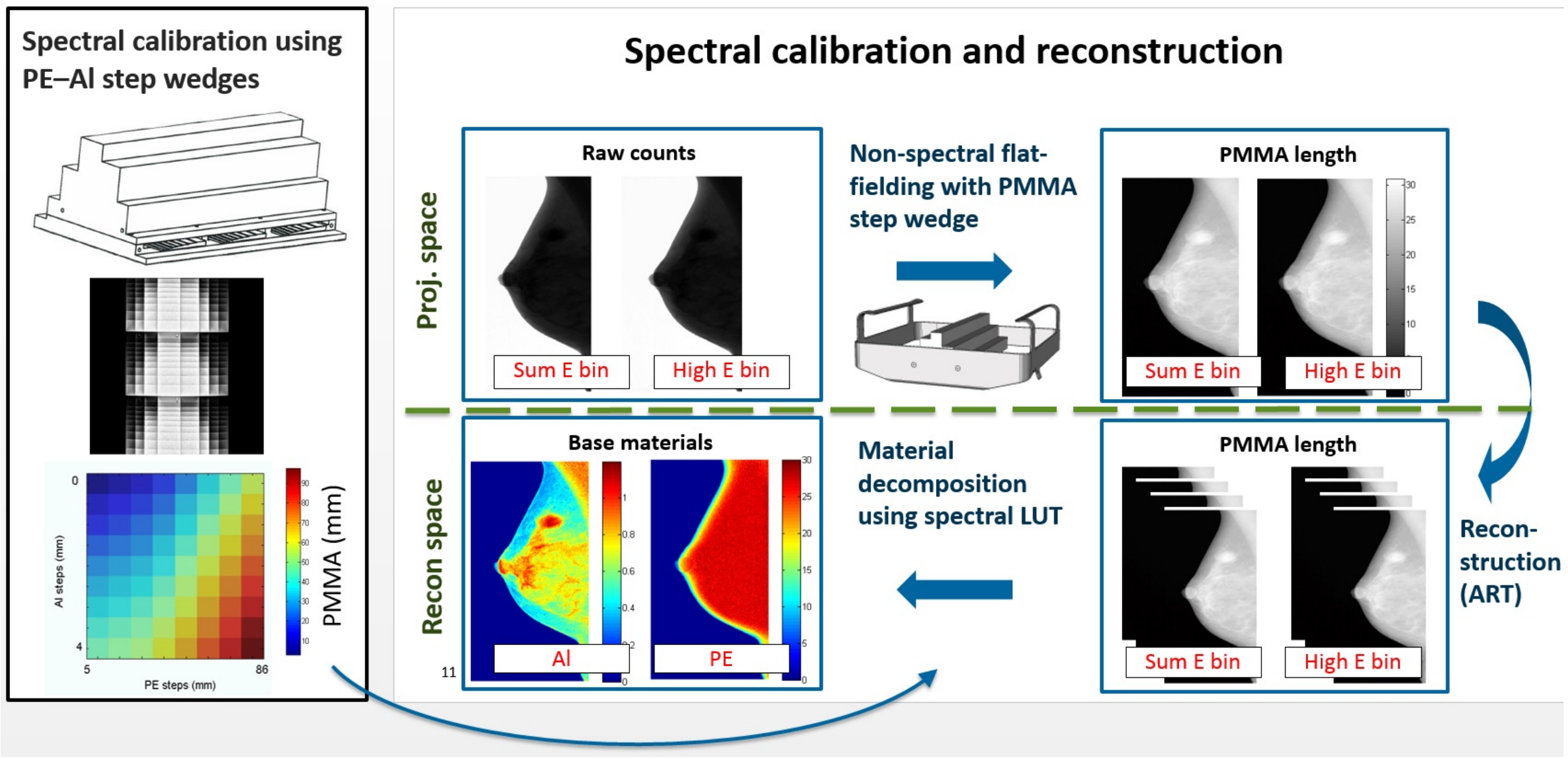

Figure 2.

## 2.3 Spectral lesion characterization

For most natural body constituents at mammographic x-ray energies, it is fair to ignore absorption edges.[6,7] X-ray attenuation is then made up of only two interaction effects, namely photoelectric absorption and scattering processes. Assuming known incident spectrum ($q_0$,$\Phi(E)$) and known detector response for the two energy bins ($\Gamma_{\mathrm{lo}}(E)$ and $\Gamma_{\mathrm{hi}}(E)$), acquisitions over two different energy ranges yield a non-linear system of equations with a unique solution for two different material thicknesses $d_1$ and $d_2$ with known linear attenuation coefficients ($\mu_1$, $\mu_2$):

$$n_{\mathrm{lo}} = q_0 \int \Phi dE \qquad (1)$$

Measurements at more than two energies yield an over-determined system of equations under the assumption of only two independent interaction processes, and would, in principle, be redundant.

As it is only possible to discriminate between two different materials in spectral imaging, the composition of overlapping (adipose and glandular) tissue needs to be known in order to discriminate between cystic and solid lesions. In 2D, the breast thickness and glandular fraction is estimated from a region surrounding the lesion and interpolated into the region of the lesion. Even though the depth information in tomosynthesis is limited and is reduced with increasing lesion size, the interpolated volume will still be smaller compared to 2D.

### 2.4 Phantom measurements

A phantom experiment was carried out for the purpose of (a) assessing the 3D decomposition in terms of homogeneity and noise, and (b) to demonstrate that the use of 3D information can potentially improve lesion characterization. Two aluminum (Al) markers with identical total thickness, but one that was distributed in depth were arranged with 50 mm of tissue-equivalent material (CIRS Inc., Norfolk VA) as background tissue (Fig. 3). The difference in Al thickness was chosen such that the two markers would simulated a cyst and a tumor, respectively. Note that we need further assumptions in order for this to be the case. If we assume homogeneous breast tissue of 50% glandularity and disregard the skin, the case with 250 $\mu$m Al represents a 9.3-mm thick cyst in a 50.39 mm thick breast, whereas the case with 200 $\mu$m Al represents an 8.7-mm thick tumor in a 49.88 mm thick breast.

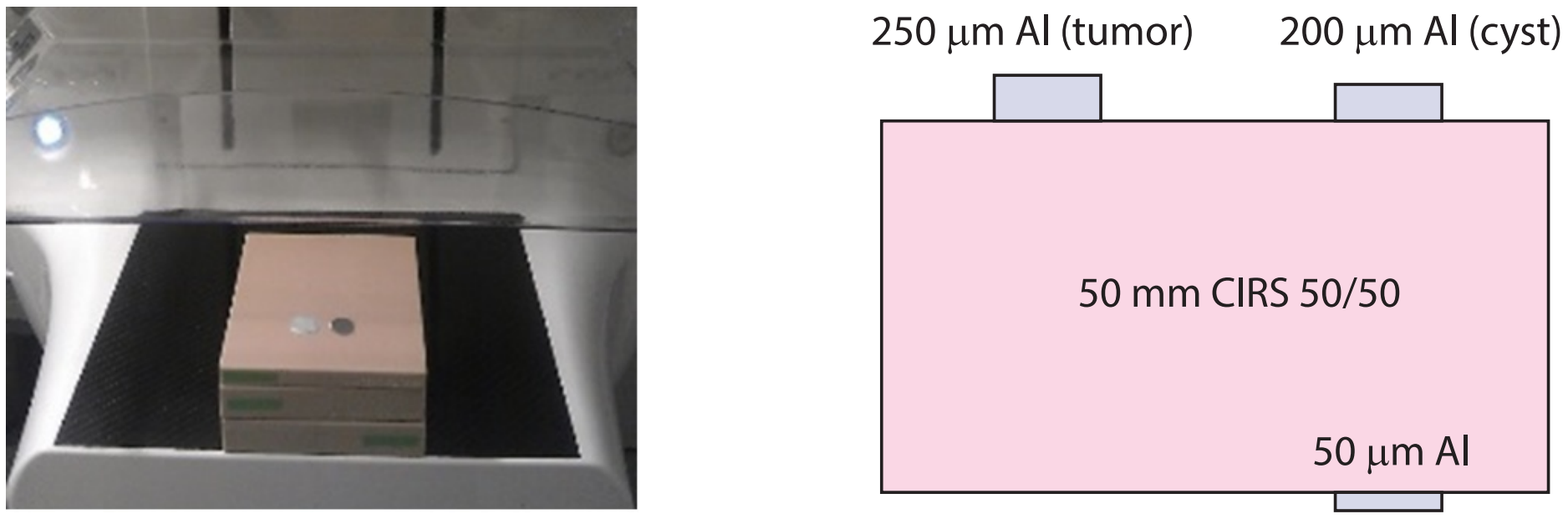

Figure 3. Schematic (left) and photograph (right) of the phantom experiment. Two 250-m Al markers arranged with a 50-mm slab of tissue-equivalent material as background tissue. The right marker is distributed in depth so that 200 m Al is on top of the tissue-equivalent material and 50 m is below the material.

### 2.5 Clinical measurements

Data from a clinical study of photon-counting spectral tomosynthesis at ImageRive, Geneva, Switzerland that was conducted early 2016 recently became available for this study. Symptomatic patients were examined with a MicroDose S0 system. All patients were asked to provide written informed consent prior to the examination and the study was approved by SwissEthics. An initial analysis of the data is presented here and we plan to extend the analysis to the final version of the study to be presented at the conference.

## 3. RESULTS

### 3.1 Phantom measurements

Figure 4 shows results of the phantom experiment in 2D, i.e. as a sum over the slices. The thickness difference between the two targets in the Al image was 15 m. The difference in thickness was likely caused by the slight misalignment between the 200 m Al on top of the phantom and the 50 m Al below the phantom, which is revealed by the conventional x-ray image.

Figure 5 shows results in 3D as a function of slice. The left target exhibits a stronger signal at the top of the bottom of the phantom. The spread in the depth direction depends on target size, but as the targets in this case are of equal size, the spread is equal and the Al thickness was normalized to the known 250 m in the left target to generate the plot on the right-hand side with quantitative values on Al thickness. The thickness of the left target was 200 m in the top slice, which is the expected value. The target thicknesses in the bottom slice are slightly higher than expected because the point-spread functions of the top targets extend through the entire volume and interfere with the measurement. The difference between the left and the right target was 30 m, which is slightly less than the expected 50 m. Figure 6 shows results from an example case in the clinical data set with a solid lesion (phyllodes tumor). The size of the lesion is similar in size as the Al target in the phantom experiment and the spread through the slices is similar in appearance.

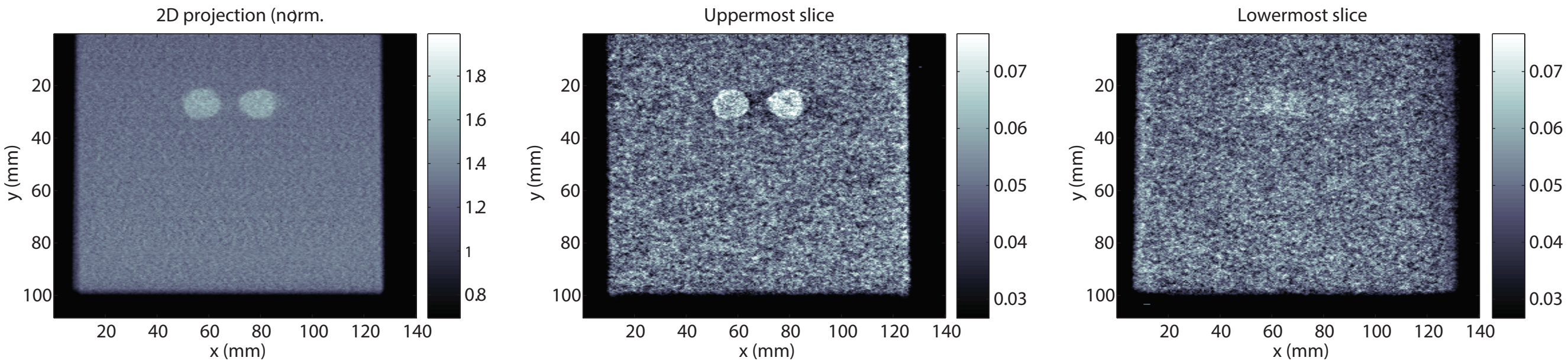

Figure 4.

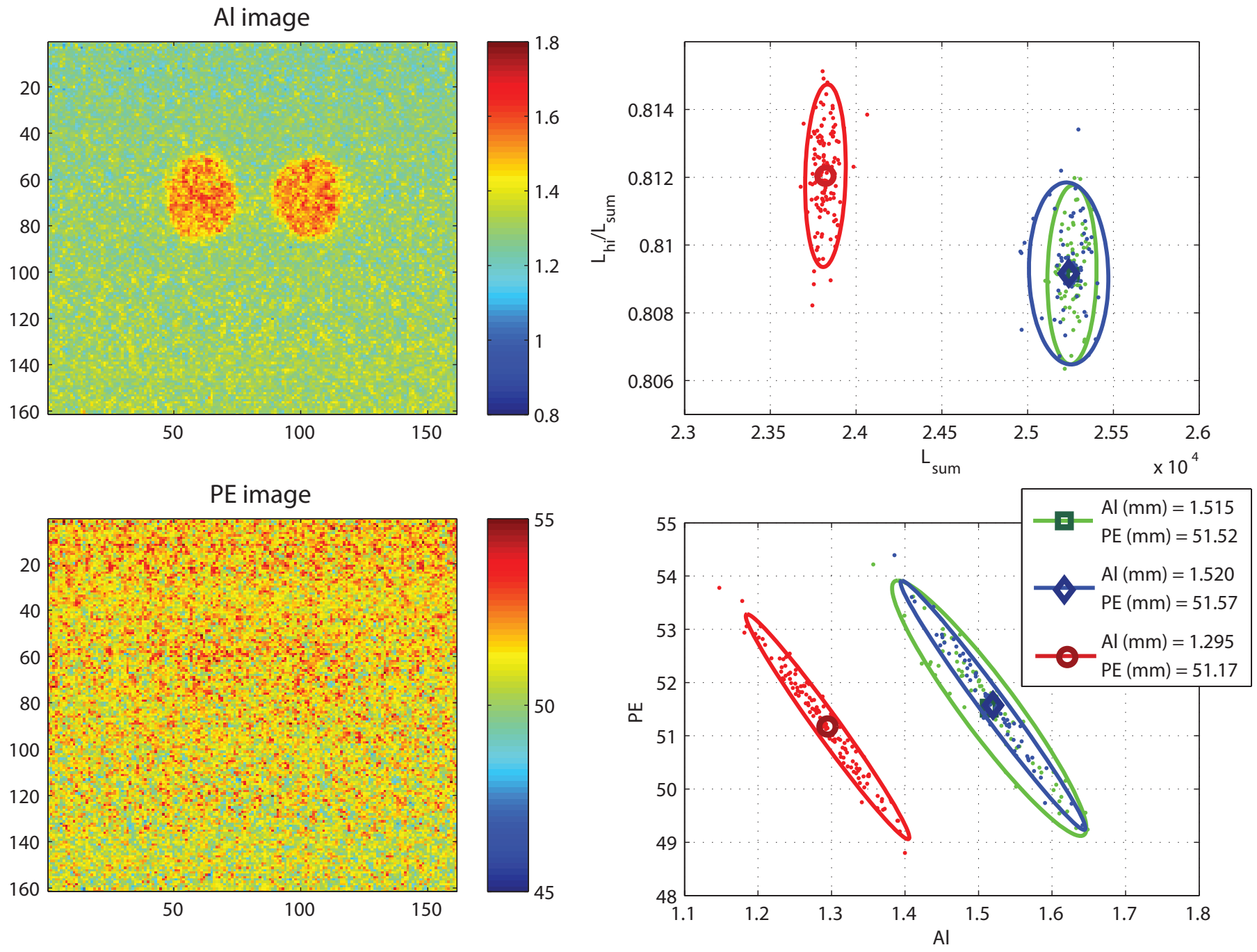

Figure 5.

### 3.2 Clinical measurements

The blue line in Figure 6, right shows the expected combination of Al and PMMA for a solid lesion of any thickness derived from specimen measurements [5]. The spectral tomosynthesis volume was decomposed into equivalent Al and PMMA thicknesses, and the lesion was evaluated in 2D (all 32 slices) and in 3D slice 27 that runs through the lesion. The 3D evaluation yields a result that is closer to the expected attenuation. It should be noted that these results are based on only one case and are so far uncertain. A more thorough investigation of a larger number of clinical cases is planned for the final study.

## 4. DISCUSSION

## 5. CONCLUSIONS

Phantom experiments showed that spectral lesion characterization is improved by 3D information from tomosynthesis. A preliminary investigation of clinical spectral tomosynthesis data also indicated an improvement when including 3D information. Improved lesion characterization holds the potential for reduction of unnecessary recalls, which would benefit individual women and the society as a whole.

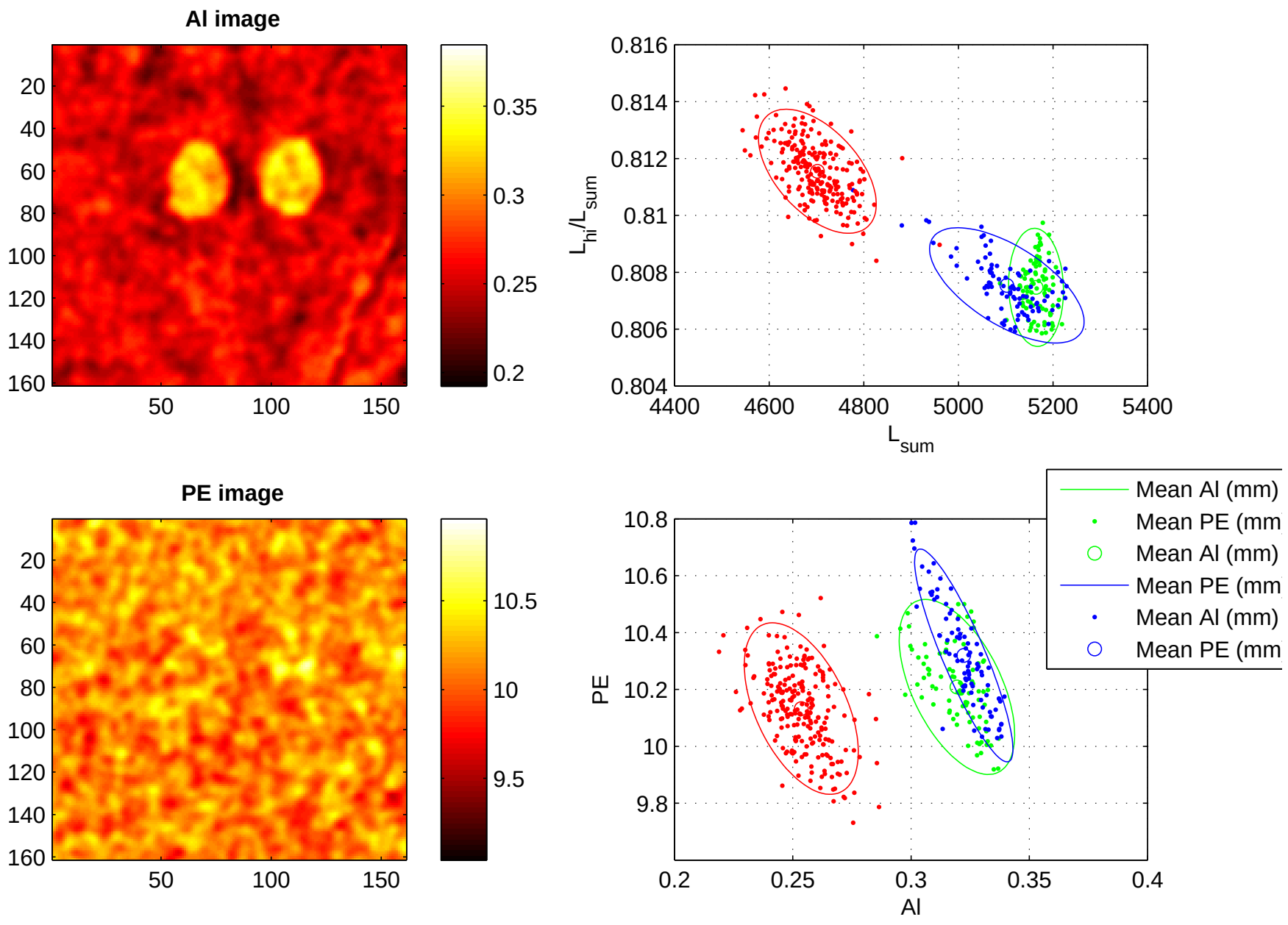

Figure 6.

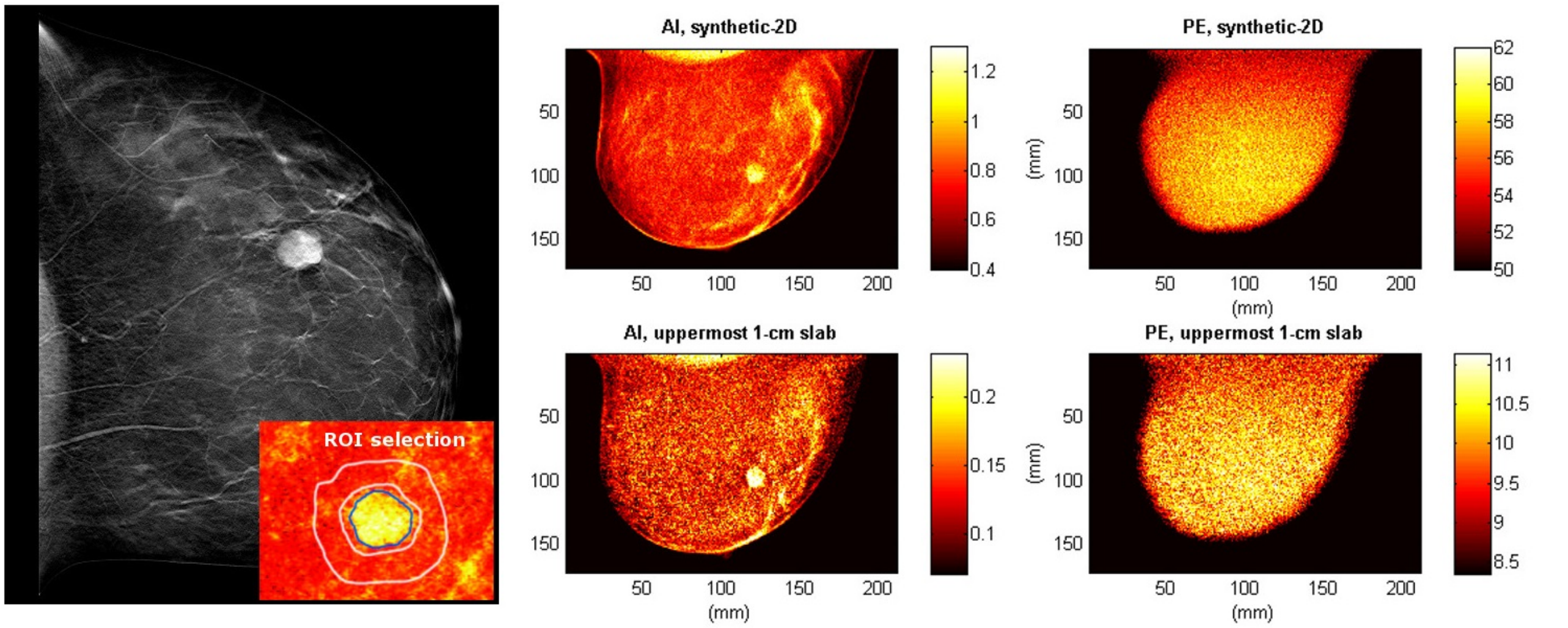

Figure 7.

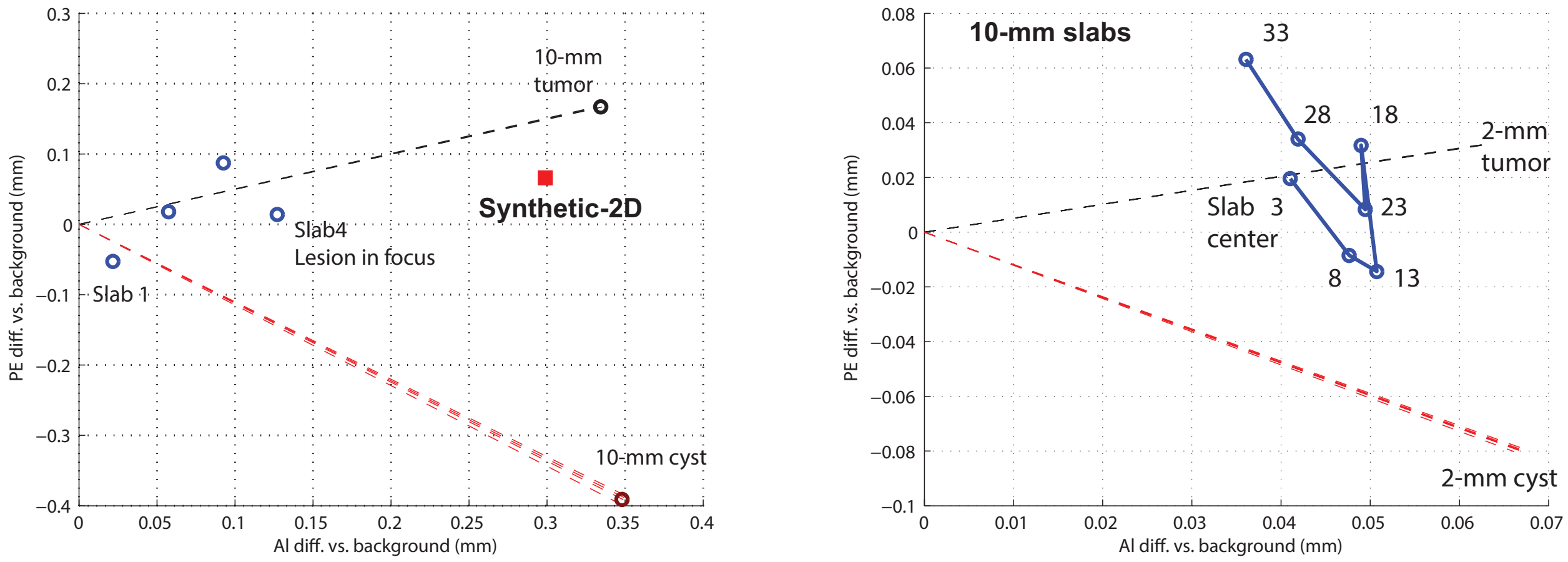

Figure 8.